\documentclass{article}
\usepackage{amsmath,graphicx}
\usepackage{amsfonts}

\date{}
\begin{document}
\title{GRAPH SIGNAL REPRESENTATION WITH WASSERSTEIN BARYCENTERS}
%
\author{Effrosyni Simou and Pascal Frossard}
%
%
%

%
\maketitle
\begin{abstract}
In many applications signals reside on the vertices of weighted graphs. Thus, there is the need to learn low dimensional representations for graph signals that will allow for data analysis and interpretation. Existing unsupervised dimensionality reduction methods for graph signals have focused on dictionary learning. In these works the graph is taken into consideration by imposing a structure or a parametrization on the dictionary and the signals are represented as linear combinations of the atoms in the dictionary. However, the assumption that graph signals can be represented using linear combinations of atoms is not always appropriate. In this paper we propose a novel representation framework based on non-linear and geometry-aware combinations of graph signals by leveraging the mathematical theory of Optimal Transport. We represent graph signals as Wasserstein barycenters and demonstrate through our experiments the potential of our proposed framework for low-dimensional graph signal representation.
\end{abstract}
%
%
\section{Introduction}
\label{sec:intro}

In many applications signals can be naturally represented on graphs. For instance, in social, transportation or brain networks. Graph Signal Processing \cite{shuman2013emerging} aims to process such signals by taking into account the underlying graph structure. The reason for this is that the connections of the graph reveal important information about the relationships between the nodes. Therefore, processing a signal defined on a graph of $N$ nodes yields more intuitive results than processing its vectorized $N$-dimensional counterpart which ignores the underlying graph structure. Given a set of graph signals, it is interesting to identify the underlying processes. This problem is addressed by dimensionality reduction methods specifically developped for graph signals. 

Unsupervised dimensionality reduction methods for graph signals have focused on dictionary learning methods where the graph is taken into consideration by imposing a structure on an overcomplete dictionary. The signals are then represented as linear combinations of a small number of atoms in the dictionary \cite{zhang2012learning}, \cite{thanou2014learning}, \cite{thanou2015multi}, \cite{yankelevsky2016dual}, \cite{yankelevsky2018dictionary}, \cite{yankelevsky2018finding}. In addition, in \cite{NIPS2013_5046} a linear wavelet transform is proposed, where the wavelets are learned with an autoencoder in order to be adaptive to a given class of signals. In practice, however, graph signals cannot always be represented using linear combinations of features. 

As a motivating example of the case where linear combinations of features may be ineffective, consider the case where the dataset may contain different instances of a diffusion process. If the dictionary contains atoms with two very distinct instances of the diffusion process, it is impossible to represent a signal at an intermediate time instance as a linear combination of those two atoms. Furthermore, increasing the number of atoms to include more instances of the diffusion process will not offer a good solution either, since the dictionary will then have a high coherence thus leading to decreased performance of the sparse coding step. 

In this work we propose a representation framework with non-linear and geometry-aware combinations of graph signal features by leveraging the mathematical theory of Optimal Transport (OT) \cite{villani2008optimal}. OT is a powerful mathematical theory that allows to define distances which exploit the geometry of the underlying space. It has thus found numerous applications, for example in image processing \cite{ferradans2014regularized}, \cite{Schmitz2018WassersteinDL}, domain adaptation \cite{courty2014domain}, traffic congestion control \cite{beckmann1952continuous}, minimum-cost network flow \cite{essid2018quadratically}, \cite{leonard2016lazy} and graphics \cite{solomon2014earth}, \cite{solomon2015convolutional}.  In our present work, we unite Optimal Transport and Graph Signal Processing with the goal to further exploit the underlying structure in graph signal representation problems. By taking into account the graph through the shortest path distance between the nodes, we employ Wasserstein distances between graph signals and compute a Wasserstein barycenter \cite{agueh2011barycenters} of a set of graph signals. We show that Wasserstein barycenters provide a geometry-aware representation of a set of graph signals. Through our experiments we demonstrate the superiority of Wasserstein barycenters compared to linear combinations for low-dimensional graph signal representation. To the best of our knowledge, our work is the first to exploit the benefits of the geometry awareness that Optimal Transport has to offer in order to enhance graph signal representation methods. 

\section{Optimal Transport Framework}
\label{sec:otframework}
Optimal Transport \cite{kolouri2017optimal} is a mathematical theory that allows to compute distances between probability measures defined on a space by leveraging the distance metric on that space. 

The OT problem was formulated by Monge \cite{monge1781memoire}. Given two probability distributions with densities $I_0, I_1 \geq 0$, $\int I_0 = \int I_1 = 1$ and their respective domains $\Omega_0$ and $\Omega_1$, the objective is to find a map $f: \Omega_0 \rightarrow \Omega_1$ that pushes $I_0$ to $I_1$ in the most efficient way with respect to a metric-dependent cost $c:\Omega_0 \times \Omega_1 \rightarrow \mathbb{R}^+$ and a mass preservation constraint $MP$. The Monge formulation of the OT problem is as follows:
\begin{equation}\label{eq:Monge}
M(I_0,I_1)=\inf_{f \in MP} \int_{\Omega_0}c(x,f(x))I_0(x)dx
\end{equation}
where $MP$ is the mass-preservation constraint which for a subset $B \subset \Omega_1$ can be expressed as: 
\begin{equation}\label{eq:MongeMassPreservation}
\int_{{x: f(x) \in B}}I_0(x)dx = \int_{B} I_1(y)dy
\end{equation}
In some cases there is no transport map $f$ that can rearrange $I_0$ to $I_1$. 
The relaxation proposed by Kantorovich \cite{kantorovich1942translation} made the solution of the transportation problem tractable in that case by allowing for split of mass. The Kantorovich OT problem aims to find a transport plan $\gamma$, where $\gamma(x,y)$ describes the amount of mass moving from $x$ to $y$, as follows:
\begin{equation}\label{eq:Kantorovich}
K(I_0,I_1)=\min_{\gamma \in MP}\int_{\Omega_0 \times \Omega_1} c(x,y)d\gamma(x,y)
\end{equation}
where the mass preservation constraint ($MP$) for $A \subseteq \Omega_0$ and $B \subseteq \Omega_1$ can now be expressed as:
\begin{equation}\label{eq:KantorovichMassPreservation}
\gamma(\Omega_0 \times B) = I_1(B) \text{ and } \gamma(A \times \Omega_1) = I_0(A)
\end{equation}

In the case where mass can only be found at specific positions, probability vectors (histograms) can be considered and the metric-dependent cost is in the form of an $n \times m$ matrix $C$, where $n$ is the dimensionality of the source histogram and $m$ is the dimensionality of the target histogram. The Kantorovich OT problem then aims to find an optimal coupling $\Gamma$ that will minimize the cost of the mass transportation with respect to the matrix $C$. The Kantorovich formulation of the OT problem for histograms is as follows:
\begin{equation}\label{eq:KantorovichHistograms}
K(\mathbf{I_0},\mathbf{I_1}) = \min_{\Gamma \in MP} \langle C, \Gamma \rangle
\end{equation}
where the mass preservation constraint ($MP$) now reads:
\begin{equation}\label{eq:KantorovichHistogramsMassPreservation}
{\Gamma \in \mathbb{R}_+^{n \times m}}: \Gamma \mathbf{1}_m=\mathbf{I_0}  \text{ and } \Gamma^{\top}\mathbf{1}_n= \mathbf{I_1}
\end{equation}
where $\mathbf{I_0}$ is the $n$-dimensional source histogram and $\mathbf{I_1}$ is the $m$-dimensional target histogram.

If $n=m$, $D$ is a distance  on $[\![ n ]\!]$ and the cost matrix $C$ in Eq. (\ref{eq:KantorovichHistograms}) can be expressed as $C = D^p$ for  $p \geq 1$, then:
\begin{equation}\label{eq:WassersteinDistanceDefinition}
W_p(\mathbf{I}_0,\mathbf{I}_1)=\left (\min_{\Gamma \in MP} \langle D^p, \Gamma \rangle \right )^{1/p}
\end{equation} 
defines the $p$-Wasserstein distance on the simplex $\Sigma_n=\{\mathbf{a} \in \mathbb{R}_+^n: \sum_{i=1}^{n}\mathbf{a}_i=1\}$.
\section{Wasserstein Distance Between Graph Signals}
\label{sec:W1forGSRep}

We now propose to use the OT framework to compare graph signals. We consider a connected, weighted and undirected graph $\mathcal{G}$ with no self-loops and non-negative weights. The set of $N$ vertices of $\mathcal{G}$ is denoted as $\mathcal{V}$ and the set of its edges as $\mathcal{E}$. The connectivity pattern of the graph is expressed via its weight matrix $\mathcal{W}$. The element $\mathcal{W}(i,j)$ corresponds to the weight of the edge connecting vertices $i,j \in \mathcal{V}$. 

We assume that signal intensities (mass) can only be found at the $N$ vertices of the graph and therefore we consider graph signals to be histograms. Thus, all considered graph signals belong to the simplex $\Sigma_N$, where $N$ is the number of vertices. Since we consider graph signals to be probability vectors, we only deal here with signals with non-negative values and normalize with the $l_1$ norm.


 
The geodesic distance of a graph captures the shortest path distance between vertices. If $s_{ij}$ is the shortest path between two nodes $i$ and $j$, the geodesic distance between them is defined as the sum of the weights of the edges in $s_{ij}$:
\begin{equation}\label{eq:GeodesicDistance}
D_G(i,j) =  \sum_{(k,l) \in s_{ij}}\mathcal{W}(k,l)
\end{equation}

We now show that $D_G$ satisfies the properties of a distance so that we can use it to define $p$-Wasserstein distances between graph signals with Eq. (\ref{eq:WassersteinDistanceDefinition}):

\begin{itemize}
\item
{ \bf Non-negativity:} Since the graph is connected, there exists a path between any two pairs of nodes $i,j$. Therefore, $D_G(i,j)>0, \forall i,j \in \mathcal{V}, i \neq j$. Furthermore, since there are no self-loops in the graph, it holds that $D_G(i,i)=0$. Therefore $D_G$ satisfies the condition of non-negativity.
\vspace*{-2.3mm}
\item
{\bf Symmetry:} Since the considered graph $\mathcal{G}$ is undirected, the shortest path from node $i$ to node $j$ is the shortest path from $j$ to $i$ reversed. As a result $D_G(i,j)=D_G(j,i)$ and therefore, $D_G$ is symmetric.
\vspace*{-2.3mm}
\item
{\bf Identity of indiscernibles:} Since the considered graph is connected with no self-loops it holds that $D_G(i,j)=0$ if and only if $i=j$. Therefore, $D_G$ satisfies the identity of indiscernibles.
\vspace*{-2.3mm}
\item
{\bf Triangle inequality:} Let $i,j,k \in \mathcal{V}$. If the shortest path from $i$ to $k$ passes through node $j$, then $D_G(i,k) = D_G(i,j)+D_G(j,k)$. If the shortest path from $i$ to $k$ does not pass through $j$, then $D_G(i,k) < D_G(i,j)+D_G(j,k)$. Therefore, $D_G$ satisfies the triangle inequality $D_G(i,k) \leq D_G(i,j)+D_G(j,k)$.
\end{itemize}

%

The computational complexity of Wasserstein distances is $O(N^3 log N)$, where $N$ is the number of nodes. Although this is not a particularly high complexity, it can become prohibitive in the case where we want to compute distances between many graph signals. As this is the case for representation learning problems, in this work we use the entropy regularized Kantorovich problem proposed by M. Cuturi in \cite{cuturi2013sinkhorn}. This problem produces ``smoothed'' approximations of the transport plan $\Gamma$ and can be solved efficiently in $O(N^2)$ with Sinkhorn projections. Furthermore, in our work we chose $p=1$ because in that case the cost for the transportation of mass is equal to the geodesic distance. The $W_1$ distance between two $\delta$ functions located at vertices $i$ and $j$ is equal to $D_G(i,j)$ and, as a result, $W_1$ distances naturally quantify translations of graph kernels throughout the graph. Thus, we compute the distance between graph signals as:

\begin{equation}\label{eq:EntropyRegularizedMetricBetweenGraphSignals}
W^{\alpha}_{1}(\mathbf{I_0},\mathbf{I_1}) = \min_{\Gamma \in MP} \langle D_G, \Gamma \rangle + \alpha H(\Gamma)
\end{equation}
where $MP$ is the mass preservation constraint as in Eq. (\ref{eq:KantorovichHistogramsMassPreservation}), $H(\Gamma)= \sum_{i,j}\Gamma_{ij}log(\Gamma_{ij}-1)$ is the negative entropy of the transport plan $\Gamma$ and $\alpha$ is the regularization parameter.

\section{Wasserstein barycenters for graph signal representation}
\label{sec:barycenters}
Equiped with the above definition of distances we now propose the use of Wasserstein barycenters \cite{agueh2011barycenters} for the representation of graph signals. 
Given a set of signals, their barycenter can be thought of as their representative mean when using Wasserstein distances. 
Thus, given a set of $M$ graph signal features $\{s_i\}_{i=1}^M$, we represent a graph signal $x$ as the Wasserstein barycenter of $\{s_i\}_{i=1}^M$:

\begin{equation}\label{eq:GraphSignalWassersteinBarycenter}
\hat{x} = \operatorname*{argmin}_{u \in \Sigma_N}\sum_{i=1}^{M} \lambda_i W_{1}^{\alpha}(s_i,u)
\end{equation}
where $W_{1}^{\alpha}$ is the entropy regularized $W_1$ distance between graph signals as defined in Eq. (\ref{eq:EntropyRegularizedMetricBetweenGraphSignals}) and $\lambda_i$ is the weight corresponding to $s_i$. The intuition behind the use of Wasserstein barycenters for the representation of graph signals as in Eq. (\ref{eq:GraphSignalWassersteinBarycenter}) is that the underlying graph structure is accounted for through the use of the $W_1^{\alpha}$ distances defined in Eq. (\ref{eq:EntropyRegularizedMetricBetweenGraphSignals}). Therefore, we propose a geometry-aware combination of the composing features of the graph signal $x$.


We construct a regular ring graph and a sensor graph of $N=64$ nodes. The sensor graph is created by placing randomly sensors in the $xy$-plane within the $[0,1]\times[0,1]$ square. The graph is created with an RBF kernel and hence the elements of the adjacency matrix are $W(i,j) = exp(- \frac{d_{ij}^2}{2\sigma^2})$, where $d_{ij}$ is the distance between the nodes $i$ and $j$. 

\begin{figure}[t!]
  \begin{minipage}[b]{1.0\linewidth}
  \begin{center}
    \includegraphics[width=8.0cm]{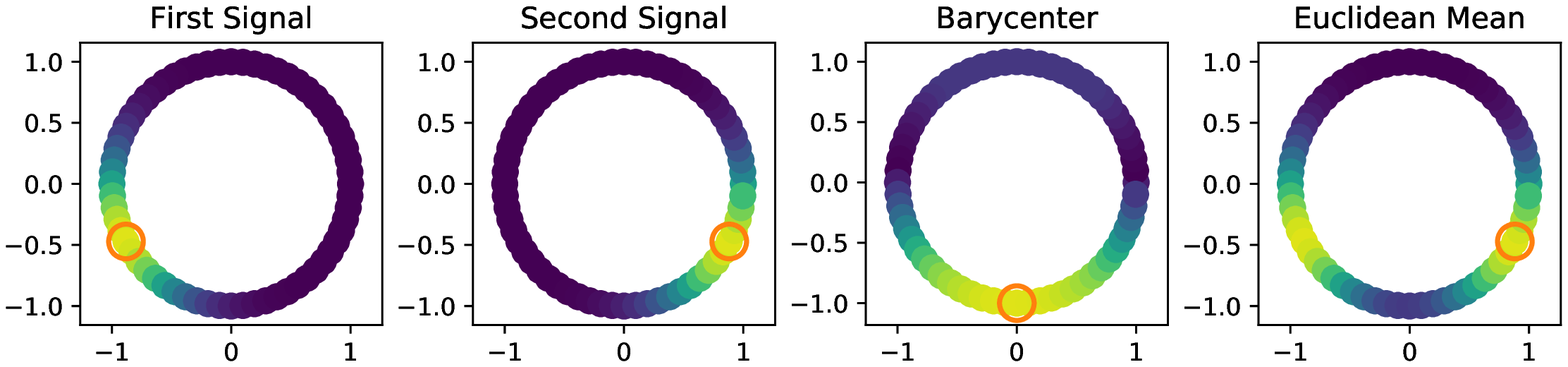}
    \end{center}
    \vspace*{-3mm}
    \centerline{(a) Ring graph.}
    \label{fig:translationRing}
  \end{minipage}
 
  \begin{minipage}[b]{1.0\linewidth}
  \begin{center}
  \includegraphics[width=8.0cm]{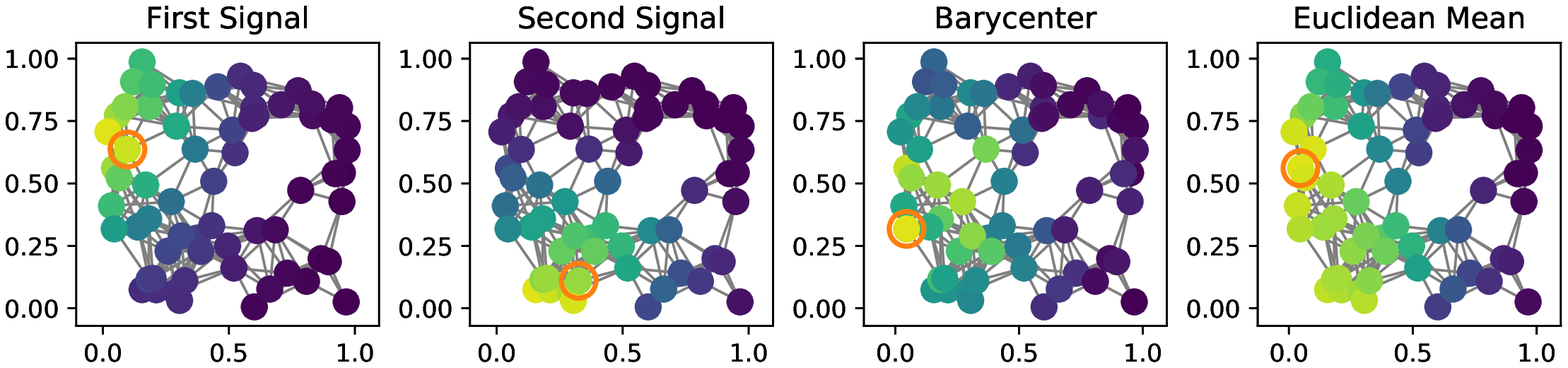}
  \end{center}
  \vspace*{-3mm}
    \centerline{(b) Sensor graph.}
  \end{minipage}
  \caption{The localized heat kernels, their $W_1^{\alpha}$ isobarycenter and their Euclidean mean. The orange circle indicates the node where the kernel is localized for the two signals and the node with the highest signal value for the barycenter and the Euclidean mean.}
  \label{fig:translation}
\end{figure}

We first consider a low-pass heat kernel ($\hat{g}_{\tau}(\lambda_l)=e^{-\tau\lambda_l}$) and we localize it at two distinct nodes of the graph. We then compute the entropy regularized barycenter of these two kernels as expressed in Eq. (\ref{eq:GraphSignalWassersteinBarycenter}) for $\alpha=0.001$. Because of the entropic regularization, the barycenter is expected to be more ``spread out" than the initial signals, as explained in Section \ref{sec:W1forGSRep}. The two signals, their barycenter and their Euclidean mean are shown in Fig. \ref{fig:translation}. It can be observed that the $W_1^{\alpha}$ barycenter is a translated localized heat kernel similar to the initial signals, albeit smoothed because of the entropic regularization. The Wasserstein distances between the graph kernels capture the geometry of the underlying domain and the barycenter is also a heat kernel positioned between the two initial signals. The Wasserstein barycenter aims  to find the most similar graph signal to the initial signals both in terms of the type of the graph signal and the position on the topology. On the contrary, Euclidean distances are completely ignorant of the underlying space and result to average representations of the graph signals that carry no geometric interpretation.

\begin{figure}[t!]
  \begin{minipage}[b]{1.0\linewidth}
  \begin{center}
    \includegraphics[width=8.0cm]{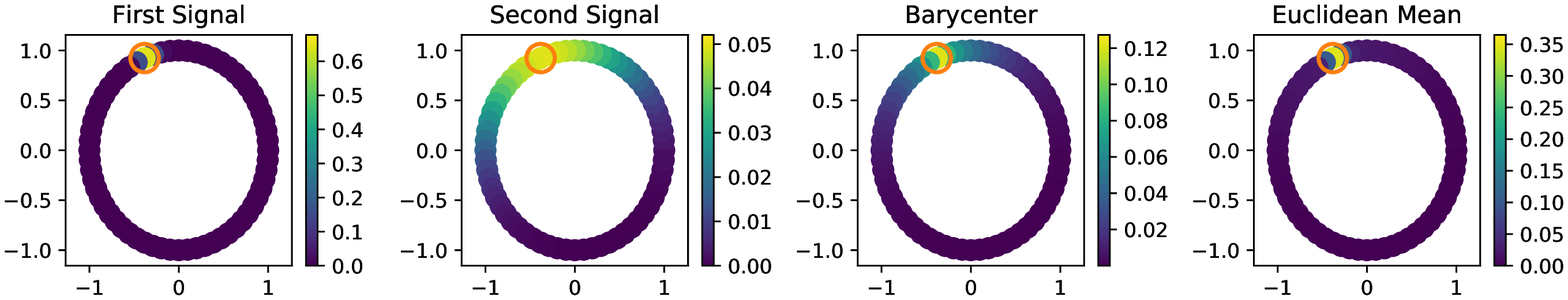}
    \end{center}
    \vspace*{-3mm}
    \centerline{(a) Ring graph.}
  \end{minipage}
 
  \begin{minipage}[b]{1.0\linewidth}
  \begin{center}
  \includegraphics[width=8.0cm]{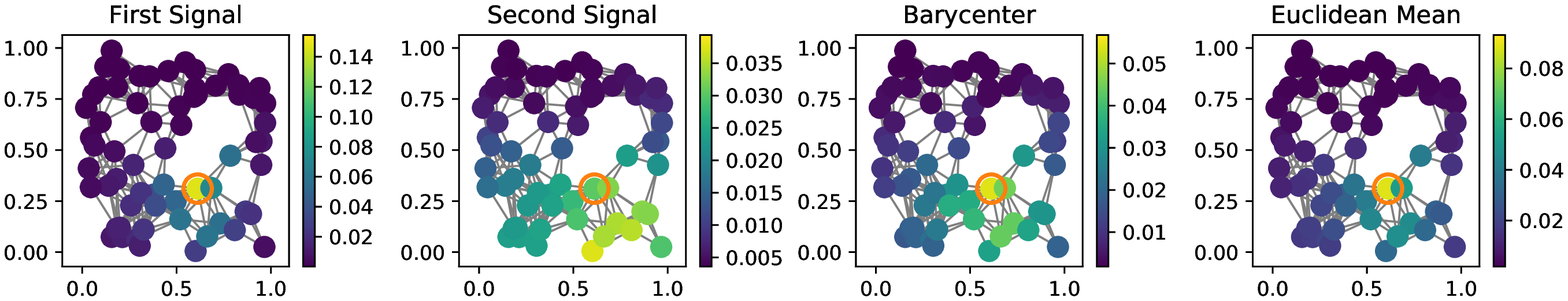}
  \end{center}
  \vspace*{-3mm}
    \centerline{(b) Sensor graph.}
  \end{minipage}
  \caption{The two instances of the heat diffusion process, their $W_1^{\alpha}$ isobarycenter and their Euclidean mean. The orange circle indicates the node where the heat diffusion process starts.}
  \label{fig:diffusion}
\end{figure}

\begin{figure}[b!]
  \begin{minipage}[b]{1.0\linewidth}
  \begin{center}
    \includegraphics[width=8.0cm]{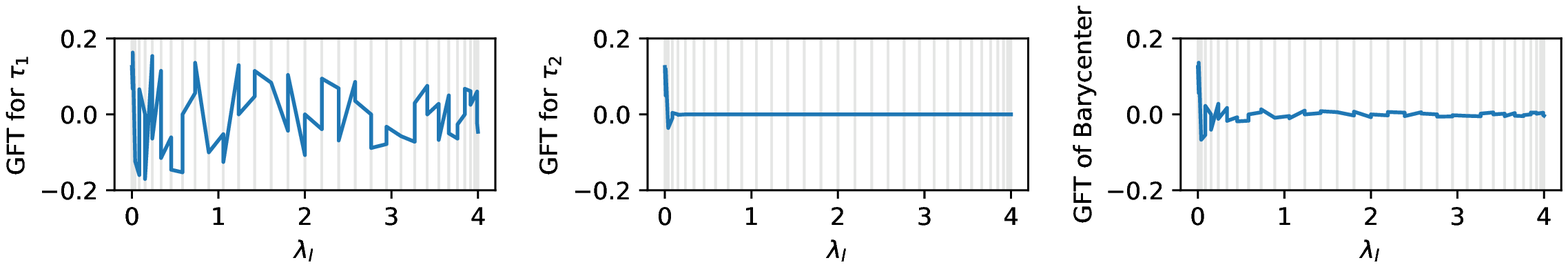}
    \end{center}
    \vspace*{-6mm}
    \centerline{(a) Ring graph.}
  \end{minipage}
 
  \begin{minipage}[b]{1.0\linewidth}
  \begin{center}
  \includegraphics[width=8.0cm]{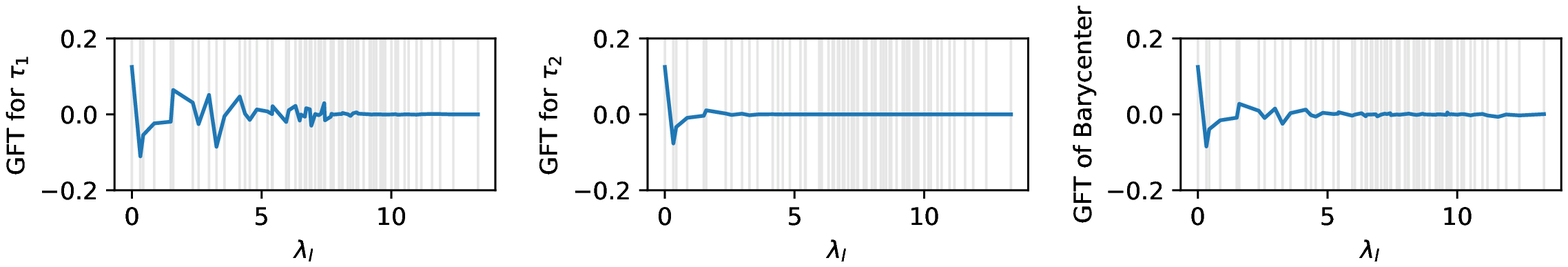}
  \end{center}
  \vspace*{-6mm}
    \centerline{(b) Sensor graph.}
  \end{minipage}
  \caption{Graph Fourier Transforms (GFT) of the two instances of the heat diffusion process and their $W_1^{\alpha}$ barycenter.}
  \label{fig:spectral}
\end{figure}

We now consider two instances of a heat diffusion process starting at a node. The first is early on in the diffusion process ($\tau_1$) and the second one is later ($\tau_2$). We compute the entropy regularized $W_1^{\alpha}$ barycenter with $\alpha=0.01$. The two graph signals, their $W_1^{\alpha}$ barycenter and their Euclidean mean are shown in Fig. \ref{fig:diffusion}. It can be observed that the $W_1^{\alpha}$ barycenter resembles to a heat diffusion process at an intermediate time instance $\tau$ ($\tau_1 \leq \tau \leq \tau_2$). This can be verified in the spectral domain, as can be seen in Fig. \ref{fig:spectral}. For the first time instance $\tau_1$ the signal is very localized and therefore its spectrum has components in almost all the frequencies. For the time instance $\tau_2$ the process has diffused in most of the graph and therefore it is a smooth signal whose spectrum only contains information in the low frequencies. We would expect the diffusion process at an intermediate time step  $\tau$ to have a spectrum that has more significant high frequency components than the second signal. This is the case for the $W_1^{\alpha}$ barycenter. In contrast, the Euclidean mean does not produce geometrically interpretable results, as can be seen in Fig. \ref{fig:diffusion}. It simply yields the average value of the signal values at each node and, as a result, completely ignores the ``expansion" of the diffusion process through the graph. 


\section{DIMENSIONALITY REDUCTION FOR GRAPH SIGNALS WITH WASSERSTEIN BARYCENTERS}
\label{sec:DictLearning}
We now illustrate the potential of the Wasserstein barycenters introduced in Section \ref{sec:barycenters} for dimensionality reduction of graph signals. For this purpose we adapt the Wasserstein Dictionary Learning (WDL) algorithm proposed in \cite{Schmitz2018WassersteinDL} for images in order to test whether it would yield meaningful features for the case of graph signals. The WDL problem is as follows:

Given a set of $S$ signals $\{x_i\}_{i=1}^{S}$, learn the dictionary $D$ and the weights $\Lambda$ such that the difference between the signals and their representations as Wasserstein barycenters of the atoms of the dictionary is minimal according to a loss $\mathcal{L}$:

\begin{equation}\label{eq:WassersteinDictionaryLearning}
\operatorname*{min}_{D,\Lambda}\mathcal{E}(D,\Lambda):=\sum_{i=1}^S\mathcal{L}(P(D,\lambda_i),x_i)
\end{equation}
where $P(D,\lambda_i)$ is the representation of $x_i$ as the barycenter of the atoms of $D$ with weights $\lambda_i$. By substituting $P(D,\lambda_i)$ with our proposed Wasserstein barycenter representation for graph signals in Eq. (\ref{eq:GraphSignalWassersteinBarycenter}) we can employ WDL \cite{Schmitz2018WassersteinDL} on graph signals.

\begin{figure}[t!]
  \begin{minipage}[b]{0.48\linewidth}
    \includegraphics[width=4.0cm]{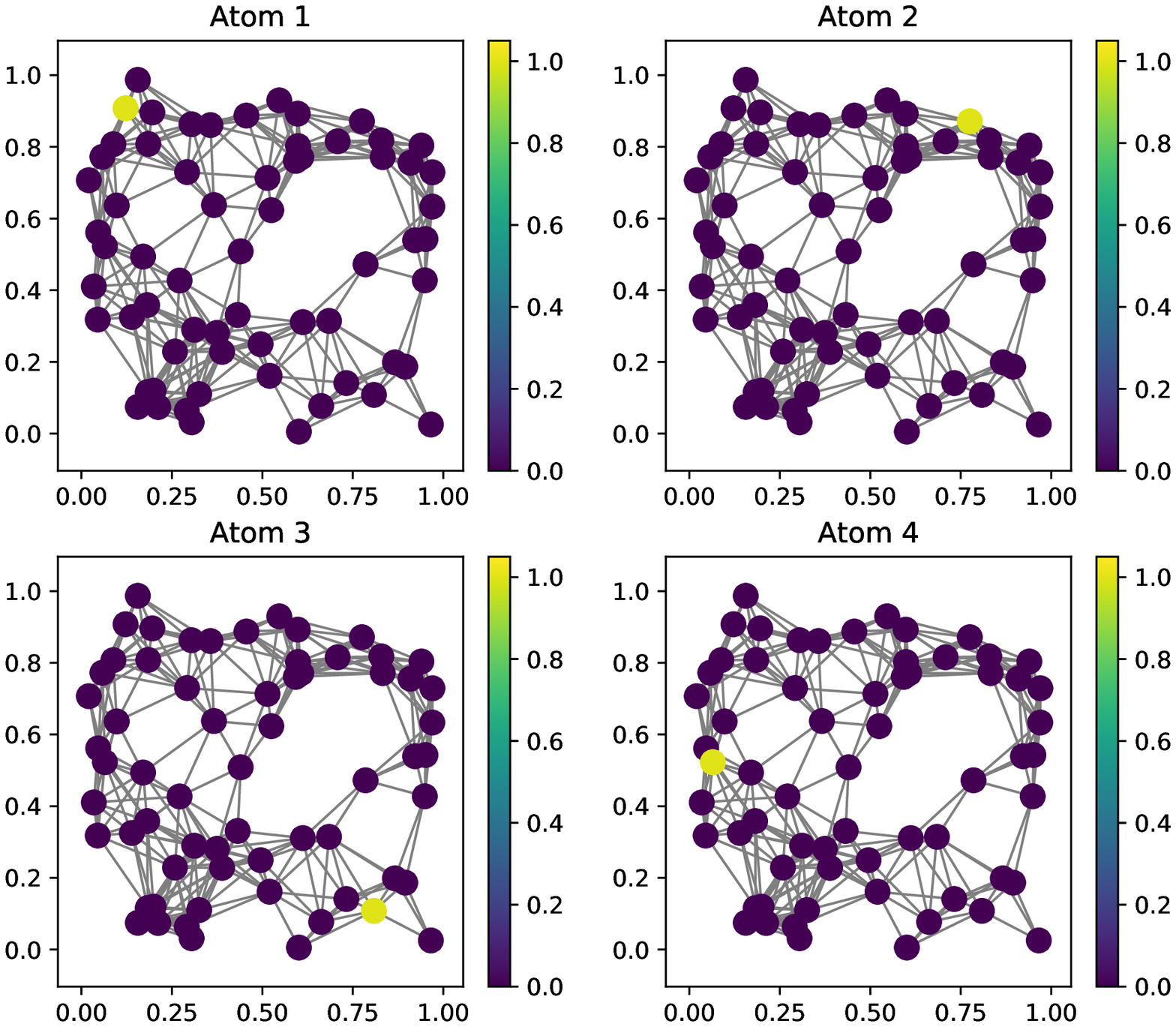}
    \centerline{(a) WDL}
  \end{minipage}
 \hfill
  \begin{minipage}[b]{0.48\linewidth}
  \includegraphics[width=4.0cm]{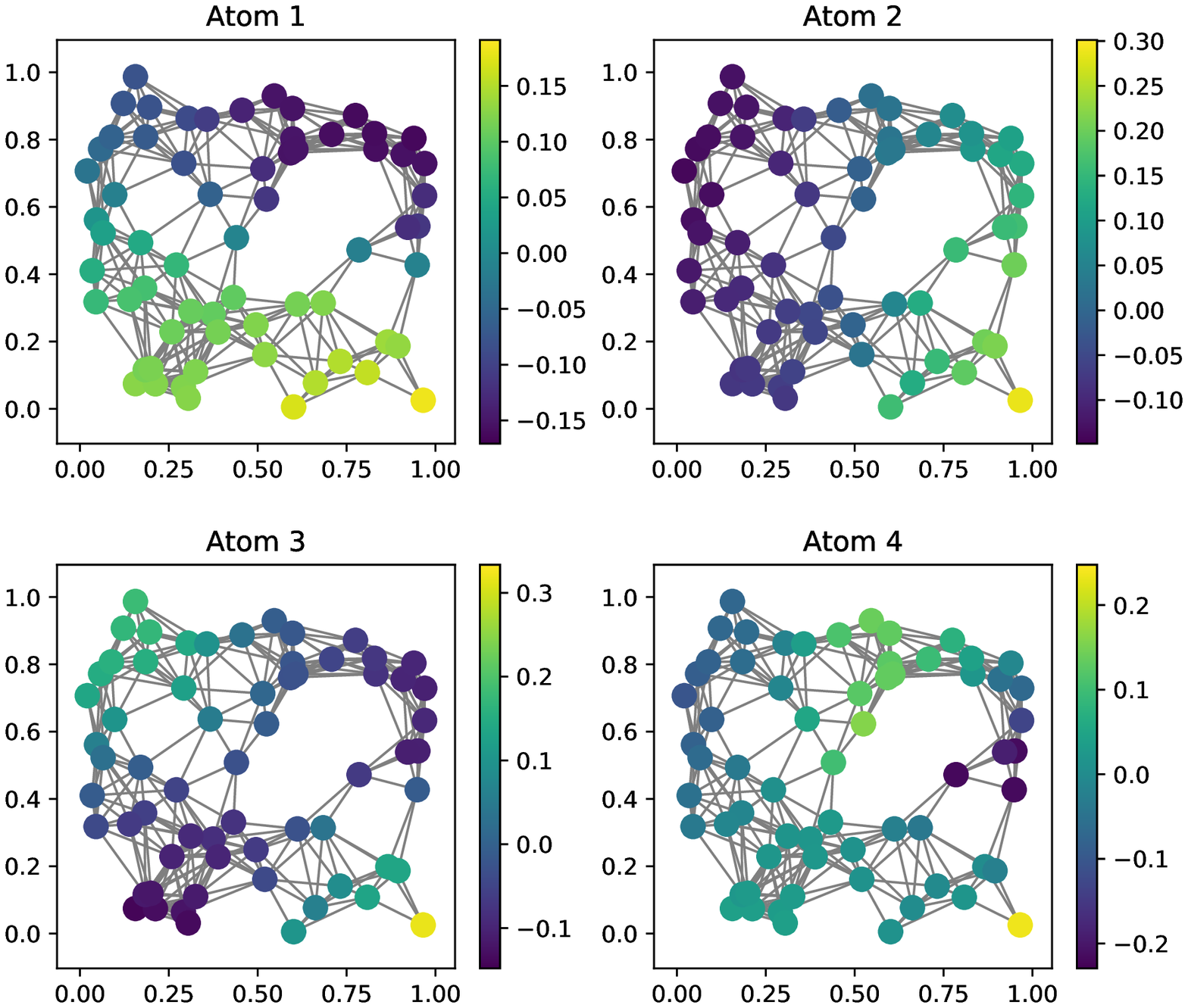}
    \centerline{(b) SVD}
  \end{minipage}
  \caption{Representations obtained with WDL and SVD.}
  \label{fig:atoms}
\end{figure}

We consider the sensor graph of Section \ref{sec:barycenters} and as training signals the graph signals obtained when we translate a localized heat kernel ($\tau=5$) at each node of the graph. 
We solve the problem in Eq. (\ref{eq:WassersteinDictionaryLearning}) with $P(D,\lambda_i)$ computed with Eq. (\ref{eq:EntropyRegularizedMetricBetweenGraphSignals}) and (\ref{eq:GraphSignalWassersteinBarycenter}), $\mathcal{L}=L_2$ and learn a dictionary of $M=4$ atoms. The atoms learned with WDL are shown in Fig. \ref{fig:atoms}a. It can be observed that the atoms learned are very localized heat kernels (almost $\delta$ functions) located at four nodes positioned at the ``edges" of the graph. This result is expected based on the intuition developed in Section \ref{sec:barycenters}. Since the entropy-regularized barycenter of two heat kernels is a smoother heat kernel positioned ``in between", it was expected when solving the inverse problem of finding the graph kernels whose entropy regularized barycenters can represent heat kernels with $\tau=5$ to recover sharper heat kernels localized at ``extreme" positions on the graph. The atoms learned truly reveal the underlying process of the translation of a heat kernel along the graph. 

Since WDL \cite{Schmitz2018WassersteinDL} was not developed for graph signals, no structure or prior is imposed on the dictionary $D$. Therefore, in order to ensure a fair comparison, we chose a dimensionality reduction method that imposes no structure on the dictionary, but uses linear combinations instead of Wasserstein barycenters. We therefore perform dimensionality reduction with Singular Value Decomposition (SVD) and obtain the representations shown in Fig.\ref{fig:atoms}b. It can be observed that although the underlying process is simply the translation of a localized heat kernel, it is not captured by the representations of SVD. This is due to the fact that in SVD the underlying graph structure is in no way accounted for. We have therefore illustrated the benefit of geometry-aware combinations with Wasserstein barycenters compared to linear combinations that ignore the underlying space.  


\section{Discussion and Future Work}
\label{sec:discussion}
In this paper we have introduced the use of Optimal Transport theory in Graph Signal Processing with a view of further exploiting the underlying graph structure in representation learning problems for graph signals. We have developed a framework for computing distances between graph signals that take into account the underlying graph and we have proposed to use Wasserstein barycenters for dimensionality reduction of graph signals. In our future work we plan to incorporate the proposed framework in representation learning algorithms developed specifically for graph signals.




\vfill\pagebreak


\bibliographystyle{IEEEbib}
\bibliography{strings,refs}

\end{document}